\documentclass[twocolumn]{aastex631}
\usepackage{graphics,graphicx}
\hypersetup{urlcolor=blue}
\usepackage{natbib}
\usepackage[outdir=./]{epstopdf}
\usepackage{textcomp,gensymb}
\usepackage{xfrac}
\usepackage{xcolor}
\usepackage{multirow}
\usepackage{longtable}

\newcommand{\kepler}{{\it Kepler}}

\newcommand\kms{km~s$^{-1}$}

\newcommand{\gaia}{{\textit Gaia}}

\usepackage[outdir=./]{epstopdf}
\usepackage{ulem}
\usepackage{mathrsfs}
\usepackage{fontawesome}
\usepackage{comment}
\usepackage[version=4]{mhchem}

\newcommand{\tess}{\textit{TESS}}
\newcommand{\ktwo}{{\textit K2}}

\newcommand{\jwst}{\textit{JWST}}

\newcommand{\starname}{TOI-2076}

\pdfoutput=1

\shorttitle{} 
\shortauthors{}

\begin{document}

\title{\tess\ Investigation - Demographics of Young Exoplanets (TI-DYE) III: an inner super-Earth in TOI-2076}

\correspondingauthor{Madyson G. Barber}
\email{madysonb@live.unc.edu}

\author[0000-0002-8399-472X]{Madyson G. Barber}
\altaffiliation{NSF Graduate Research Fellow}
\affiliation{Department of Physics and Astronomy, The University of North Carolina at Chapel Hill, Chapel Hill, NC 27599, USA} 

\author[0000-0003-3654-1602]{Andrew W. Mann}
\affiliation{Department of Physics and Astronomy, The University of North Carolina at Chapel Hill, Chapel Hill, NC 27599, USA}

\author[0000-0001-7246-5438]{Andrew Vanderburg}
\affiliation{Department of Physics and Kavli Institute for Astrophysics and Space Research, Massachusetts Institute of Technology, Cambridge, MA 02139, USA}

\author[0000-0001-6037-2971]{Andrew W. Boyle}
\altaffiliation{NSF Graduate Research Fellow}
\affiliation{Department of Physics and Astronomy, The University of North Carolina at Chapel Hill, Chapel Hill, NC 27599, USA}

\author[0009-0006-9572-1733]{Ana Isabel Lopez Murillo}
\altaffiliation{UNC Chancellor’s Science Scholar}
\affiliation{Department of Physics and Astronomy, The University of North Carolina at Chapel Hill, Chapel Hill, NC 27599, USA}

\begin{abstract}

Young ($<$500 Myr) multi-planet transiting systems are valuable environments for understanding planet evolution by offering an opportunity to make direct comparisons between planets from the same formation conditions. TOI-2076 is known to harbor three, 2.5-4 R$_\oplus$ planets on 10-35 day orbits. All three are {\it JWST} cycle 3 targets (for transmission spectroscopy). Here, we present the detection of TOI-2076 e; a smaller (1.35 R$_\oplus$), inner (3.02 day) planet in the system. We update the age of the system by analyzing the rotation periods, Lithium equivalent widths, color-magnitude diagram, and variability of likely co-moving stars, finding that TOI-2076 and co-moving planetary system TOI-1807 are $210\pm20$ Myr. The discovery of TOI-2076 e is motivation to revisit known transiting systems in search of additional planets that are now detectable with new \tess\ data and updated search methods. 

\end{abstract}


\section{Introduction} \label{sec:intro}

Population level analyses of \kepler\ and \ktwo\ transiting systems show that mature multi-planet systems are largely tightly-packed \citep{Muirhead2015, Brewer2018}, having a high level of intra-system uniformity in radius and period space that is much more similar than expected by chance \citep{Lissauer2011, Dawson2016, Weiss2018}. The current population of known young multi-planet systems do not show this same level of intra-system uniformity \citep{Thao2024}, seemingly suggesting that planet systems become uniform over time. 

However, the mature \kepler\ sample is dominated by small planets \citep[$\lesssim$4 R$_\oplus$;][]{Petigura2022}, while young planets in \tess\ are overwhelmingly large \citep[$\gtrsim$4 R$_\oplus$;][]{Fernandes2022, Vach2024}. In order to determine if intra-system uniformity is primordial or an evolutionary effect, we need to be consistently sensitive to the smallest planets in the youngest systems.  

One challenge of detecting small planets around young stars is the high levels of stellar activity present. This obscures small planet signals and can adversely affect light curve processing. However, custom light curve processing \citep[e.g.,][]{Vanderburg2019} and search methods tuned for young stars \citep[e.g.,][]{Rizzuto2017} have enabled the discovery of younger \citep[e.g.,][]{Barber2024_iras} and smaller \citep[e.g.,][]{Mann2018_zeit6} planets. HD 63433 d, an Earth-sized planet on a tight orbit around a member of the 400 Myr Ursa Major group \citep{Capistrant2024}, demonstrates how processing instrumental systematics and stellar variability simultaneously could increase our sensitivity to additional, smaller planets in known young transiting systems. This system also highlights the need for such work to study the origins of intra-system uniformity; the original two planets had similar radii, but the additional smaller planet was a mismatch in both radius and period (low intra-system uniformity). 

As part of our larger survey searching for planets in young stellar associations \citep[the TI-DYE survey;][]{Barber2024_iras, Barber2024_hipc}, we revisit known transiting systems searching for missed planets. With our increased sensitivity, we are able to recover smaller planets that could aid in understanding whether uniformity results from formative \citep[e.g.,][]{Poon2020} or evolutionary processes \citep[e.g.,][]{Lopez2013}. 

Young multi-planet systems are also found near period resonances at a much higher rate than their older counterparts \citep{Dai2024}. In turn, they are more likely to exhibit detectable transit-timing variations (TTVs; Lopez-Murillo in prep). Given the challenges of measuring radial velocities in the presence of stellar noise \citep{Blunt2023}, TTVs remain a more promising route for measuring masses and eccentricities of young planets. However, this method only works if we detect all the major perturbing planets.

Since the discovery, TOI-2076 has been observed in two additional \tess\ sectors, and we have made numerous improvements to our transit-search pipeline \citep[see][]{Barber2024_hipc}, motivating a new search. Here, we report the discovery of an inner, smaller, fourth planet in the system: TOI-2076 e. Due to different prior analyses yielding different ages for TOI-2076, we re-derive the age using the rotations, Lithium levels, color-magnitude diagram positions, and variabilities of identified co-moving stars. Given the systems prior TTV detection \citep[][Lopez Murillo et al. in prep]{Osborn2022} and {\it JWST} data \citep{Feinstein2024_kronos_prop}, the discovery of TOI-2076 e and updated age analysis will aid in the interpretation of existing data on the system.

In this paper, we detail the discovery and age analysis of TOI-2076\,e. We describe our light curve extraction in Section \ref{sec:lc} and our transit search in Section \ref{sec:search}. Taking into account the stellar parameters described in Section \ref{sec:stellarParams}, we derive the planet parameters in Section \ref{sec:transit_analysis}, rule out false positive scenarios in Section \ref{sec:fp}, and run an injection-recovery analysis in Section \ref{sec:injrec}. We search for co-moving stars to TOI-2076 in Section \ref{sec:selectGroup} and use these members to determine an age of the cluster in Section \ref{sec:Age}. We discuss potential additional known planets in the cluster in Section \ref{sec:knownPlanets} and search for additional transiting systems in Section \ref{sec:additionalSearch}. Finally, we discuss the impact of this new planet and age analysis in Section \ref{sec:summary}.

\section{\tess\ Light Curve}\label{sec:lc}

\begin{figure*}
    \centering
    \includegraphics[width=0.98\linewidth]{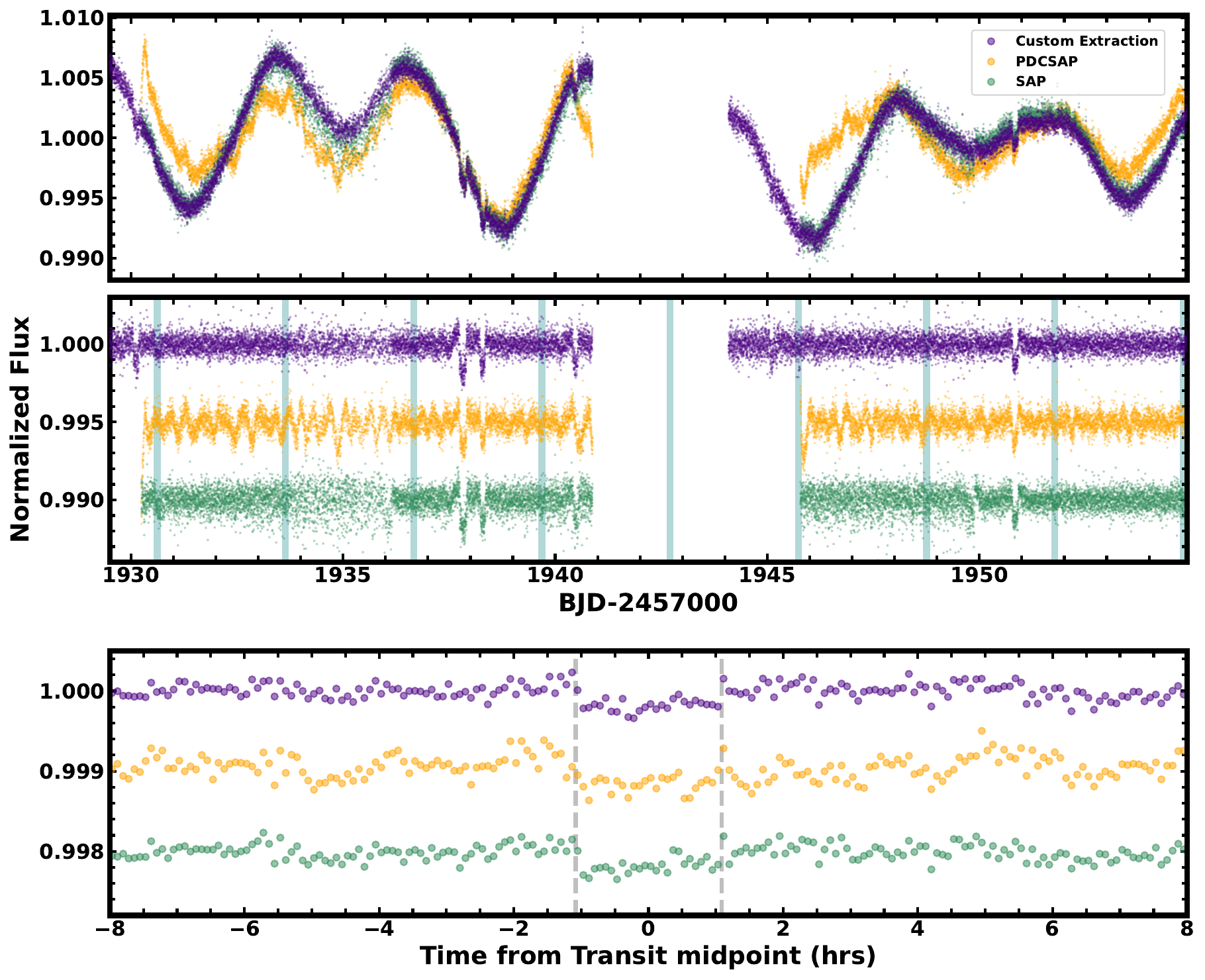}
    \caption{Our custom-extracted light curve (purple) compared to the PDCSAP (orange) and SAP (green) light curves for Sector 23. The top panel shows the raw, normalized light curves, and the middle panel shows the \texttt{notch}-detrended light curves, using a 0.75 day filtering window, with the transits of TOI 2076 e highlighted in teal and a y-offset between the extractions for clarity. The raw and detrended custom light curves show significantly less noise than the PDCSAP light curves. The bottom panel shows the phase-folded light curves binned to 5-minute intervals with a y-offset for clarity. The gray dashed lines mark the expected ingress and egress. The custom-extraction light curve shows a clear transit-like shape while the PDCSAP light curve is too noisy to detect the transit. }
    \label{fig:customVSpdcsap}
\end{figure*}

\starname\ (TIC 27491137) was first observed by \tess\ in Sector 16, from 2019 September 12 to 2019 October 6, and re-observed in Sector 23 (2020 March 19 - 2020 April 15), Sector 50 (2022 March 26 - 2022 April 22), and Sector 77 (2024 March 25 - 2024 April 23). The target was pre-selected for 2-minute short-cadence light curves for Sectors 16 and 23 and 20-second short-cadence light curves for Sectors 50 and 77. During Sector 77, \tess\ was put into Safe Mode April 6 - April 16 due to the on-board data recorder filling up due to a missing datalink on April 1. Thus, Sector 77 contains only 17.92 days of science data\footnote{\url{https://archive.stsci.edu/missions/tess/doc/tess_drn/tess_sector_77_drn108_v01.pdf}}. All \tess\ light curves used in this analysis can be found in MAST: \dataset[10.17909/t9-st5g-3177]{http://dx.doi.org/10.17909/t9-st5g-3177} and \dataset[10.17909/t9-nmc8-f686]{http://dx.doi.org/10.17909/t9-nmc8-f686}.

\subsection{Extraction Pipeline} \label{sec:extraction}

For our analysis, we used a custom light curve extraction pipeline starting from the Simple Aperture Photometry \citep[SAP;][]{Twicken2010SAP, Morris2020SPOC} from the Science Processing Operations Center \citep[SPOC;][]{Jenkins2016SPOC}, utilizing the shortest cadence available for each sector. Our systematic corrections generally followed \citet{Vanderburg2019}. To summarize, we corrected the SPOC SAP light curves using a linear model consisting of a basis spline with 0.2 day breaks to model low-frequency variations, several moments of the distributions of the spacecraft quaternion time series measurements within each exposure, seven co-trending basis vectors from the SPOC Presearch Data Conditioning \citep[PDC;][]{Smith2012, Stumpe2012, Stumpe2014} band-3 flux time series correction with the largest eigenvalues, and a 0.1 day high-pass time series from the SPOC background aperture.

We estimated the uncertainties on the flux by taking the median of three values: 1) the median absolute deviation of the point minus the adjacent point; 2) the median absolute deviation of the flattened light curve \cite[flattened using \texttt{lightkurve};][with a 4$\sigma$ outlier threshold]{lightkurve}; and 3) sigma clipping, applying a median filter, sigma clipping again, and fitting a Gaussian to the resulting distribution of points. Each method was broadly consistent with each other, and the final results are not dependent on the uncertainty estimate. Estimating the uncertainties for each sector separately, and we adopted an uncertainty of 0.0005 for sectors 19 and 23 and 0.001 for sectors 50 and 77.

\section{Transit search and identification of the planet}\label{sec:search}

Our ongoing search for transiting planets in young stellar associations \citep{Barber2024_iras, Barber2024_hipc} includes a search for additional transiting planets in known young systems. This program takes advantage of better light curve processing (Section~\ref{sec:extraction}) and improvements to the transit-search (below). \starname\ was one of the first targets where we identified additional transit-like signals. 

We used the updated \texttt{Notch and LOCoR} \citep[\texttt{N\&L;}][]{Rizzuto2017} pipeline as described in \cite{Barber2024_iras}. Using a 0.75-day filtering window, \texttt{Notch} detrended the light curve using a second-order polynomial while preserving trapezoidal, transit-like signals. At each point, \texttt{notch} calculates the change in the Bayesian Information Criterion (BIC) based on the improvement when adding the trapezoid to the model compared to the plain polynomial. We then searched the BIC time-series for periodic signals between 0.5 and 50 days with an SNR $>$ 8 using a box-least squares (BLS) search. We recovered a 10.4 day signal (TOI-2076 b) with a BLS SNR of 33, a 21.0 day signal (TOI-2076 c) with a BLS SNR of 32, a 35.1 day signal (TOI-2076 d) with a BLS SNR of 56, and a 3.0 day signal (TOI-2076 e) with a BLS SNR of 17.

In Figure \ref{fig:customVSpdcsap}, we highlight the sensitivity of the combination of the \texttt{Notch} pipeline with the custom light curve extraction; the planet was not recovered at sufficient SNR (generally $>7$) in the PDCSAP or SAP light curve and recovered only at modest SNR ($\simeq$10) using the custom extraction with a more traditional search method (a high-pass filter and BLS). 

\section{Stellar Properties}\label{sec:stellarParams}

\cite{MacDougall2023} combined photometry, high-resolution spectroscopy, and \gaia\ parallaxes to derive stellar properties of 108 \tess\ planet-hosting stars, including TOI-2076. We chose to adopt their stellar radius ($R_* = 0.7960^{+0.0286}_{-0.0217}$), stellar mass ($M_* = 0.8557^{+0.0201}_{-0.0212}$), effective temperature ($T_{eff} = 5192^{+61}_{-58}$), and metallicity ($[Fe/H] = 0.020^{+0.058}_{-0.059}$) for our transit analysis, though these values are consistent with the stellar values derived in \cite{Hedges2021}, \cite{Osborn2022}, and \cite{Damasso2024}.

The four prior studies all estimated the age of TOI-2076: $204\pm50$ Myr from \cite{Hedges2021}, $340\pm80$ Myr from \cite{Osborn2022}, $2.7^{+3.6}_{-1.9}$ Gyr from \cite{MacDougall2023} and 300$\pm$80\,Myr from \cite{Damasso2024} \citep[based on the analysis in][]{Nardiello2022}. The \cite{MacDougall2023} age is based on isochronal fitting of just the host star, which tends to give such large uncertainties. The other three ages are all based on co-moving stars. While all ages agree, the resulting age uncertainties are larger than what is routinely achieved for young associations \citep[e.g.,][]{Newton2022,Wood2023,Thao2024}. Further, inspection of the membership lists shows significant disagreement. Given the importance of this target, we opted to revisit the membership and age of the group in Section~\ref{sec:Age}.

\section{Transit analysis and planet parameters}\label{sec:transit_analysis}

\begin{figure*}
    \centering
    \includegraphics[width=0.32\linewidth]{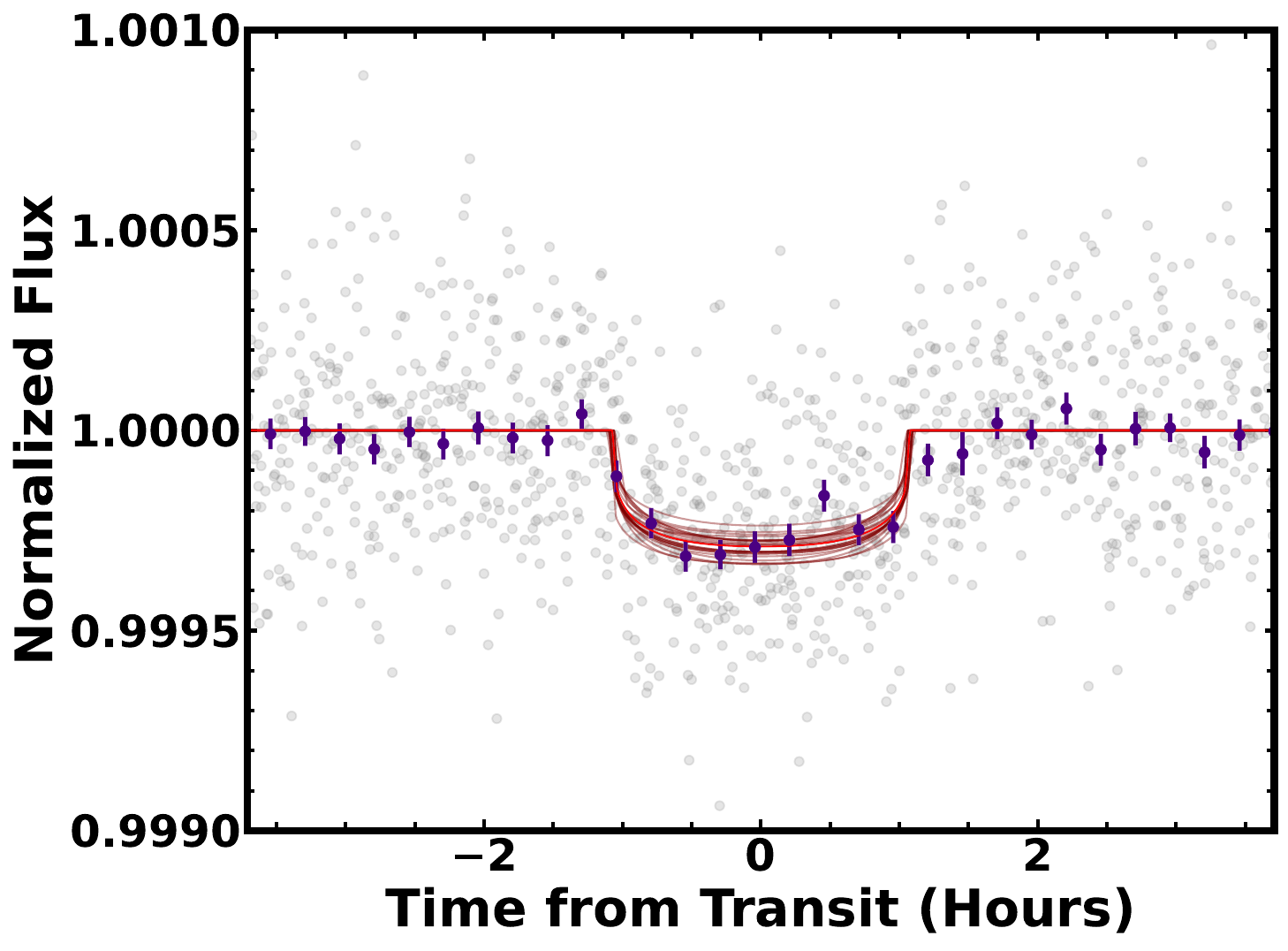}
    \includegraphics[width=.67\linewidth]{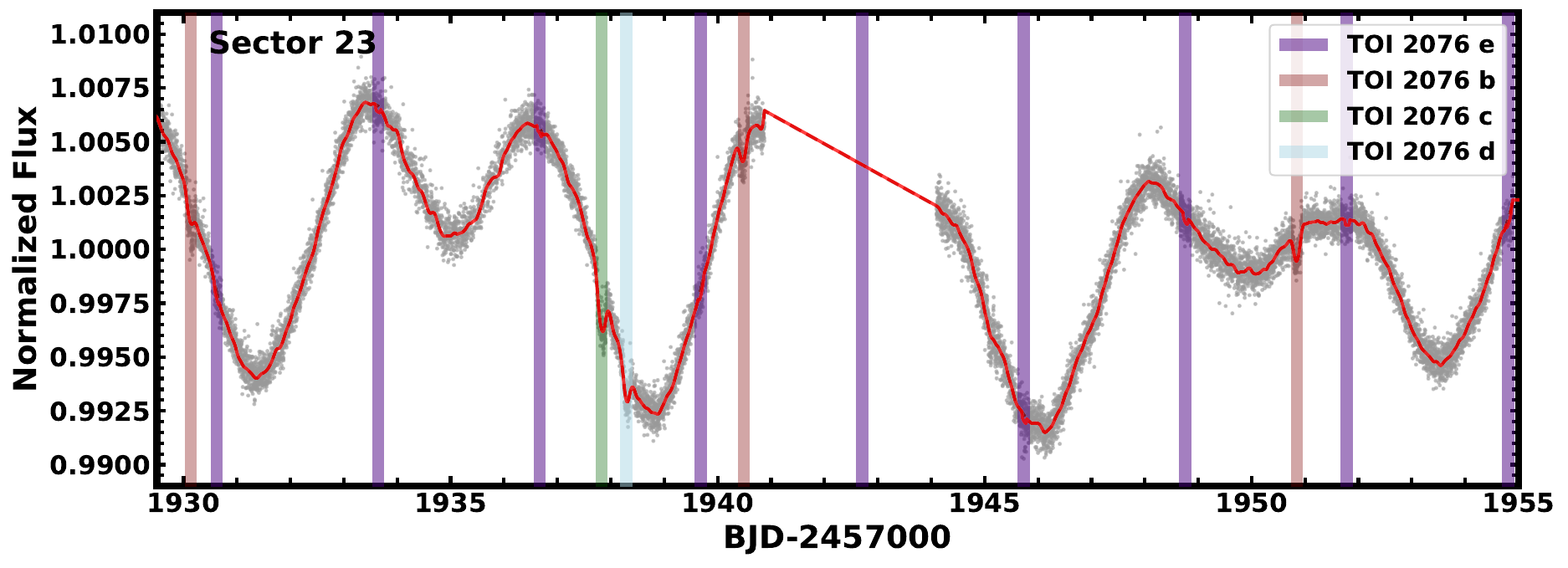}
    \caption{Left) \tess\ light curve binned to 10-minute intervals and phase-folded (gray points) with the full (un-binned) \tess\ light curve phase-folded and binned to 15-minute intervals for clarity (purple points). The best-fit transit model is shown as the bright, opaque red line with 25 model fits pulled from the posterior shown as the dark, translucent red lines. The best-fit GP stellar variability model has been removed from the data and transit model. Right) Representative section of the \tess\ light curve (gray points) with the best-fit GP model (red line). Transits of TOI 2076 e are highlighted in purple, and transits of TOI 2076 b, c, and d are highlighted in pink, green, and blue, respectively.}
    \label{fig:sector}
\end{figure*}

We fit the systematics-corrected \tess\ light curve using \texttt{MISTTBORN} \citep[MCMC Interface for Synthesis of Transits,
Tomography, Binaries, and Others of a Relevant Nature;][]{Mann2016a, MISTTBORN}\footnote{\url{https://github.com/captain-exoplanet/misttborn}}. \texttt{MISTTBORN} utilizes \texttt{BATMAN} to generate transit models \citep{BATMAN}, \texttt{celerite2} to model the stellar variability using a Gaussian Process \citep[GP;][]{celerite2}, and \texttt{emcee} to explore parameter space \citep{emcee}. 

Stochastically-driven damped simple harmonic oscillators (SHOs) have been shown to be preferred in modeling stellar variability \citep{ForemanMackey2017} and have been used extensively to model photometry of young stars with multiple transiting planets \citep[e.g.,][]{Wittrock2022, Wood2023, Thao2024}. We initially used a sum of two SHOs using the \texttt{RotationTerm} in \texttt{celerite2}. We found that the difference between the quality factors of the two oscillators was unconstrained, suggesting only a single SHO is needed. We instead fit the stellar variability using a single SHO following the description in \cite{Gilbert2022}. We find that the planetary parameters agree between the single and summed SHOs fits, but the stellar variability is better modeled by the single SHO, and we chose this model as our preferred fit.

We fit for 10 parameters to model the stellar variability and transits of TOI-2076\,e simultaneously. For the planet, we fit for the time of inferior conjunction ($T_0$), planet orbital period ($P$), planet-to-star radius ratio ($R_p/R_*$), and impact parameter ($b$). Previous fits of the system suggest the outer three planets are in near circular orbits \citep{Polanski2024}, so we restrict eccentricity to 0. We also fit for stellar density ($\rho_*$) and two quadratic limb-darkening coefficients ($q_1$ and $q_2$) following the triangular sampling prescription \citep{Kipping2013}.

The remaining three parameters were used to describe the GP stellar variability model. We fit for the undamped period of the oscillator ($P$), the damping timescale ($\tau$), and the standard deviation of the process ($\sigma$).

The parameters evolved under uniform priors with only physical limitations (see Table \ref{tab:priors}). We ran the MCMC using 50 walkers for 100,000 steps and a 20\% burn-in. The total run was more than 50 times the autocorrelation time, indicating the number of steps was sufficient for convergence. 

\begin{table}[]
    \centering
    \caption{Prior distributions used in the \texttt{MISTTBORN} \tess\ transit fit.}
    \begin{tabular}{lcc}
    \hline
    \hline
     Description  &  Parameter & Prior$^\alpha$ \\
     \hline
     \multicolumn{3}{c}{Planet Parameters}\\
     \hline
     impact parameter & $b$ & $U(0.0,1.0)$ \\
     planet-to-star radius ratio &  $R_P$/$R_*$ & $U(0.0,1.0)$ \\
     \hline
     \multicolumn{3}{c}{Stellar Parameters}\\
     \hline
     limb-darkening coefficient & $q1$ & $U(0.0,1.0)$ \\
     limb-darkening coefficient & $q2$ & $U(0.0,1.0)$ \\
     stellar density & $\rho_*$ ($\rho_\odot$) & $\rho_* > 0$ \\
     \hline
     \multicolumn{3}{c}{GP Parameters}\\
     \hline
     period & $P$ (days) & $P>0$ \\
     damping timescale & $\tau$ (days) & $\tau>P$\\
     standard deviation & $\sigma$ & $\sigma > 0$\\
     \hline
     \multicolumn{3}{l}{$\alpha$ $U(a,b)$ indicates a uniform distribution from a to b.}
    \end{tabular}
    \label{tab:priors}
\end{table}

We present the best-fit parameters in Table \ref{tab:parameters}. We also show the phased-folded light curve and a representative section of the light curve with the GP model and highlighted transit locations in Figure \ref{fig:sector}.

While we did not fit for eccentricity, we can confirm the orbit is likely near-circular from the stellar density fit. The stellar density in the transit fit is affected by transit duration, so discrepancies between the stellar density from the transit fit and the inferred stellar density from the stellar mass and radius would suggest an eccentric orbit \citep{Van-Eylen2015}. The best-fit stellar density from the transit fit ($1.65^{+0.21}_{-0.54}$ $\rho_\odot$) agrees with the inferred stellar density ($1.696\pm0.20$ $\rho_\odot$), suggesting our assumption of zero eccentricity did not impact our transit fit. 

\subsection{Transit Timing Variations}
Due to transit timing variations (TTVs) observed in the transits of TOI-2076 b, c, and d \citep[][Lopez Murillo et al. in prep]{Osborn2022}, we opted to fit only for the planet parameters of the new candidate; TOI-2076 e. The TTV semi-amplitude for b, c, and d are $\simeq10$\,minutes. Because of the shorter orbital period, we expect the semi-amplitude to be smaller in e, allowing us to use a linear ephemeris in our \tess\ transit fit.

A large TTV signal would cause the transit to smear, artificially increasing the transit duration and decreasing the transit depth. In our fit, this would result in an increased stellar density. The agreement between the best-fit stellar density and inferred density from stellar mass and radius suggests any TTV present in the transits of TOI-2076\,e has a minor, or no major, impact on our fit. 

We attempted to fit for a TTV amplitude separately from our \tess\ transit fit. Following the procedure in Lopez Murillo et al. in prep, for each transit, we fit a GP with a SHO kernel to model the local stellar variability simultaneously with a \texttt{BATMAN} transit model in an MCMC framework. We locked all transit parameters except transit midpoint (which was allowed to float within an hour of the expected midpoint) and allowed the GP parameters to float within physical boundaries, following the same priors as the \texttt{MISTTBORN} GP fit (see Table \ref{tab:priors}). We found that the individual transits were too shallow to constrain the transit timings well, with fits often not converging on a solution. We were therefore unable to constrain the TTV amplitude.

Future observations at higher SNR will be needed to check for TTV signals in the transits of TOI 2076 e.

\begin{table*} 
\centering
\caption{Parameters of TOI 2076 e} 
\begin{tabular}{lcc} 
\hline 
\hline 
Description & Parameter & Value \\ 
\hline 
\multicolumn{3}{c}{Stellar Parameters}\\
\hline
stellar density & $\rho_{\star}$ ($\rho_{\odot}$) & $1.65^{+0.21}_{-0.54}$ \\ 
limb-darkening coefficient & $q_{1}$ & $0.47 \pm 0.13$ \\ 
limb-darkening coefficient & $q_{2}$ & $0.218 \pm 0.059$ \\ 
\hline
\multicolumn{3}{c}{GP Parameters}\\
\hline 
standard deviation & $\sigma$ & $0.0214^{+0.0031}_{-0.0024}$ \\ 
period & $P$ (days) & $3.40^{+0.31}_{-0.26}$ \\ 
damping timescale & $\tau$ (days) & $3.55^{+0.39}_{-0.31}$ \\ 
\hline
\multicolumn{3}{c}{Measured Planet Parameters} \\ 
\hline 
time of inferior conjunction & $T_0$ (BJD-2457000) & $1740.21306^{+0.00081}_{-0.0008}$ \\ 
orbital period & $P$ (days) & $3.0223445^{+3.3\times10^{-6}}_{-3.2\times10^{-6}}$ \\ 
planet-to-star radius ratio & $R_P/R_{\star}$ & $0.0156 \pm 0.001$ \\ 
impact parameter & $b$ & $0.0 \pm 0.4$ \\ 
\hline 
\multicolumn{3}{c}{Derived Parameters} \\
\hline 
semi-major axis to stellar radius ratio & $a/R_{\star}$ & $10.4^{+0.43}_{-1.0}$ \\ 
inclination & $i$ ($^{\circ}$) & $90.0 \pm 2.3$ \\ 
transit duration (first to fourth contact) & $T_{14}$ (days) & $0.0902^{+0.0015}_{-0.0013}$ \\ 
planet radius & $R_P$ ($R_J$) & $0.1209^{+0.009}_{-0.0087}$ \\
& $R_P$ ($R_\oplus$) & $1.355^{+0.101}_{-0.098}$ \\
semi-major axis & $a$ (AU) & $0.0385^{+0.0021}_{-0.0049}$ \\ 
equilibrium temperature$^\dagger$ & $T_{\mathrm{eq}}$ (K) & $1138.0^{+72.0}_{-27.0}$ \\ 
\hline
\multicolumn{3}{l}{$\dagger$ assuming zero albedo}
\end{tabular} 
\label{tab:parameters}
\end{table*}

\section{False positive analysis}\label{sec:fp}

\cite{Hedges2021} used archival Palomar Observatory Sky Survey \citep{Minkowski1963, Reid1991} and PanSTARRS \citep{Chambers2016} images, new high-resolution imaging from a range of facilities, and the star's high proper motion \citep[pm = $118.27$ mas/yr;][]{GaiaCollaboration2023} to rule out the presence of a stellar companion as well as a background or foreground star. \citet{Damasso2024} gathered more than 300 spectra spanning $\simeq$3 years, which rules out any companion that would be too tight to detect in the imaging \citep{Wood2021}.

Using the contrast curves from \cite{Hedges2021} and the flattened (GP removed) light curve from \tess, we ran \texttt{TRICERATOPS} \citep{triceratops, Giacalone2021} to determine an initial false-positive probability (FPP). \texttt{TRICERATOPS} calculates the probabilities the signal is various transit-like scenarios in a Bayesian framework. We calculate a FPP $\sim$ $10^{-5}$. While this is well below the threshold for validation, it may be an underestimate of the true FPP because \texttt{TRICERATOPS} requires a flattened light curve and our GP stellar variability model is calculated assuming a transit-like signal is present. 

Further evidence supports TOI-2076 e is a real planet:
\begin{itemize}
    \item Multi-transiting systems have lower intrinsic (prior) probabilities of being a false positive \citep{Lissauer2011, Rowe2014, Valizadegan2023}.
    \item Following \cite{Vanderburg2019}, we use the transit shape and depth to calculate the faintest companion that could cause the signal to be $\Delta$T $<$ 2.5 mags. A companion of this magnitude (T$<$10.8) would have been detected in the suite of imaging or spectra  discussed above.
    \item While the transit depth is too small to analyze individual transit events, folded transits from each of the three sectors are consistent over a period of four years. Stellar signals would evolve over this period.  
\end{itemize}

Overall, the low false-positive probability based on the transit shape and the suite of follow up observations confirm that TOI-2076 e is a real planet.

\section{Injection-Recovery Analysis}\label{sec:injrec}

\begin{figure}[t]
    \centering
    \includegraphics[width=0.99\linewidth]{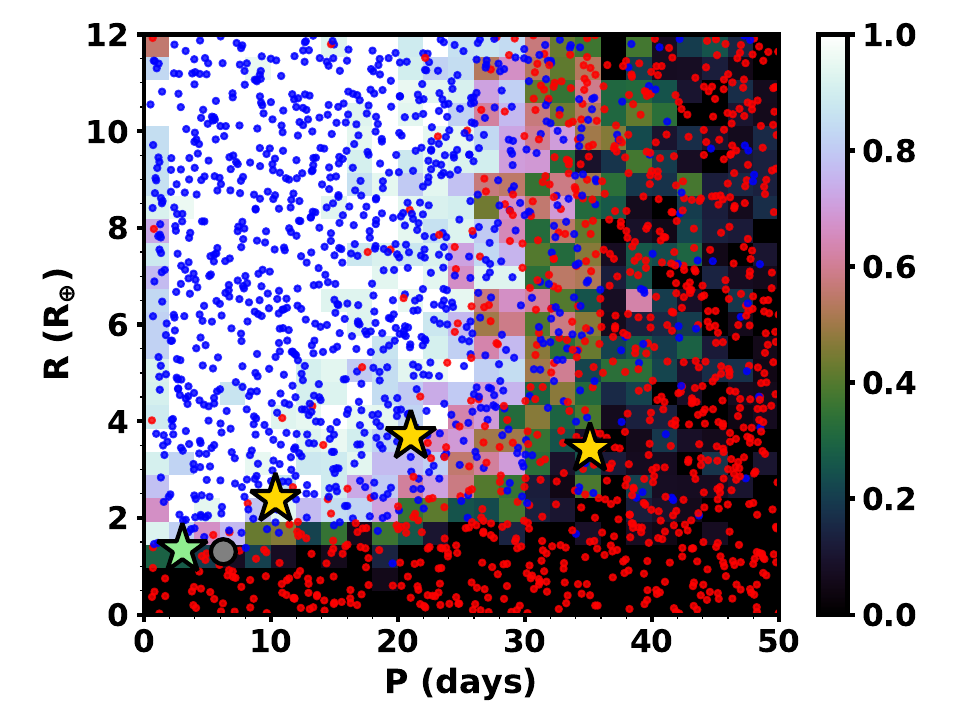}
    \caption{Period-radius injection-recovery map for TOI 2076. Blue points indicate recovered signals, while red points mark signals that were not recovered. Only 20\% of injected planets are shown for clarity. The background is color-coded by the overall completeness in a given bin. The previously known planets in the system are marked as the gold stars, TOI 2076 e is marked as the green star, and candidate TOI 2076 f is marked as the gray circle.}
    \label{fig:injrec}
\end{figure}

We preformed an injection-recovery analysis to test the detection limits of our pipeline. This is another check of the reliability of our signal (signals in low-completeness zones are more likely to be false alarms) and to determine if there may be other planets residing below our detection limits.

We injected 10,000 synthetic planet signals, following the procedure in \cite{Rizzuto2017}. We randomly generate planets with a period 0.5-50 days, radius 0.01-12 R$_\oplus$, impact parameter 0-1, and $T_0$ dependent on the period (between the beginning of the dataset and the pulled orbital period). We restrict eccentricity to 0 for simplicity. The synthetic planet signal is injected into the raw light curve (as extracted in Section \ref{sec:lc}), and we attempt to recover the planet by re-running \texttt{Notch}. We show the results in Figure \ref{fig:injrec}. TOI 2076 e falls in a parameter space with an approximately 70\% completeness rate, suggesting TOI 2076 e is not likely to be a false alarm.

\subsection{TOI 2076 f?}

The resonance chain of the system as it stands is 3.4:1-2:1-5:3. A planet with a period of approximately 6 days would complete the resonance chain: 2:1-5:3-2:1-5:3. Our pipeline detected an additional signal at 6.25 days with a $T_0$ = 2458740.686 BJD, which we initially rejected due to low SNR and from visual inspection. If real, this planet would have a radius of $\sim$1.3-1.4$R_\oplus$. Based on our injection-recovery tests (Section~\ref{sec:injrec}) completeness is close to 0\% here. Future \tess\ data or follow-up photometry are needed to confirm or refute this signal.

\section{Selection of group members}\label{sec:selectGroup}

\begin{figure*}[ht]
    \centering
    \includegraphics[width=0.98\linewidth]{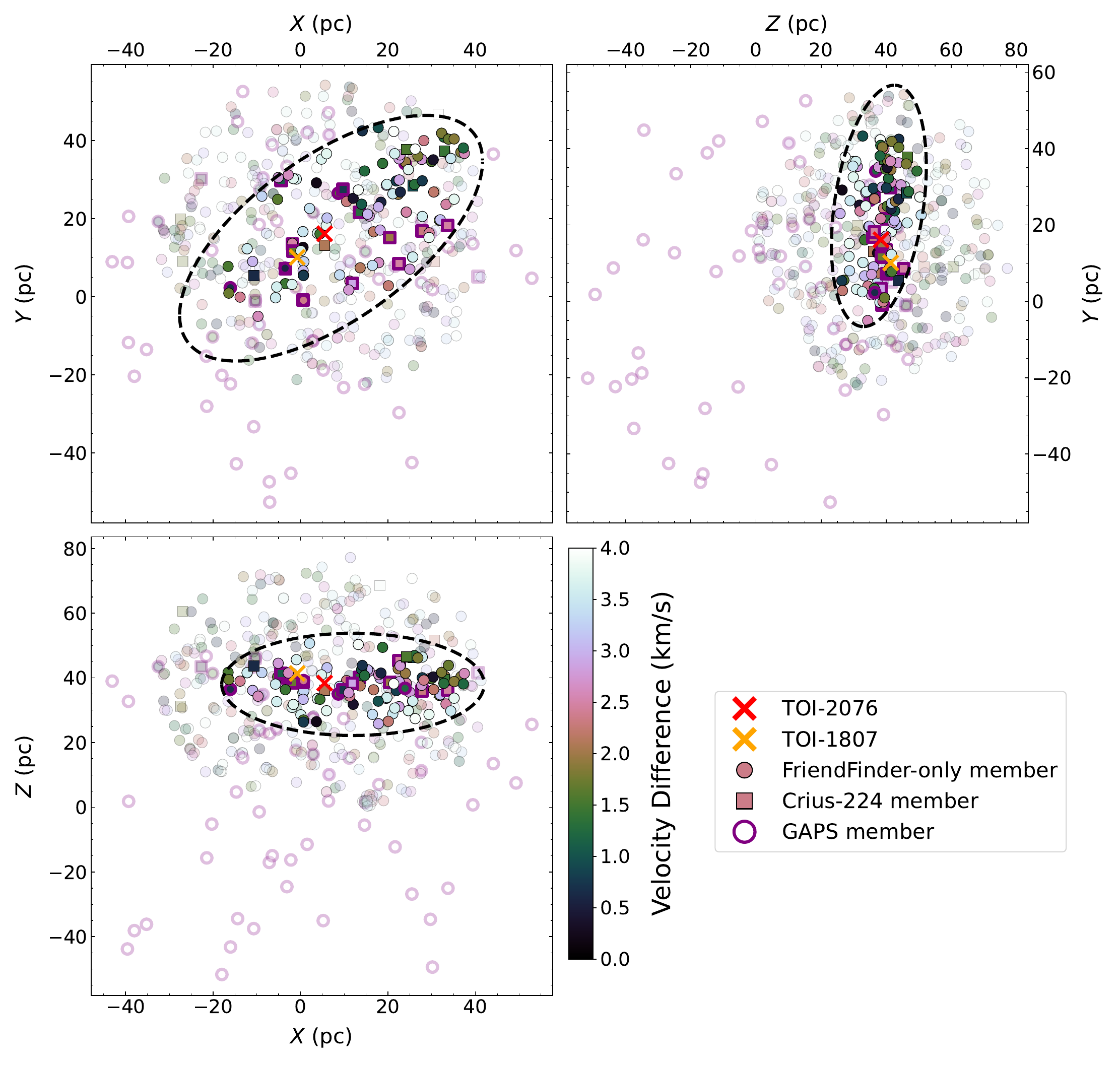}
    \caption{Candidate co-moving stars to TOI-2076 (red x) and TOI-1807 (orange x). All stars identified with \texttt{FriendFinder} are shown as colored points. Stars identified as a member of Crius 224 from \cite{Moranta2022} are marked as a square, and stars identified as a member in the GAPS survey \citep{Nardiello2022, Damasso2024} are outlined in purple. Stars that fall into all three black, dashed lined regions are used in the age analysis. Stars not used in the analysis but identified as candidate members are shown in lower opacity.}
    \label{fig:groupXYZ}
\end{figure*}

TOI-2076 has been previously clustered into groups by the Crius survey \citep[Crius 224, 0.1-0.7 Gyr;][]{Moranta2022} and by the GAPS survey \citep[$300\pm80$ Myr;][]{Nardiello2022}. These clusters, however, are small (26 and 76 members, respectively) which can impact the age determination if there are older field interlopers identified as members \citep{THYMEVI}. For simplicity, we will refer to the population as Crius 224 for the rest of the paper. 

We searched for additional co-moving stars in Crius 224 using \texttt{FriendFinder}\footnote{\url{https://github.com/adamkraus/Comove}} \citep{THYMEV}. To summarize, \texttt{FriendFinder} uses \gaia\ Data Release 3 astrometry and an input radial velocity of the star of interest to compute XYZ positions and UVW velocities of nearby stars. \texttt{FriendFinder} then looks for stars with tangential velocities and XYZ coordinates within user-set bounds of the target. 

We initially restricted our selection to stars with a tangential velocity ($V_{tan}$) within 4 \kms\ and a 3D distance within 40 pc. This resulted in a list of 450 co-moving stars, 402 of which were unique to the \texttt{FriendFinder} list. We combined all three sets of stars, \texttt{FriendFinder}, Crius 224, and the GAPS survey, to create a list of 489 candidate members.

There was a clear string of over density of nearby, co-moving stars in XYZ space but a large portion of contaminating stars in the field. Our goal is to derive a precise age for Crius 224/TOI-2076, for which we want a clean (rather than complete) list, so we applied elliptical cuts designed to encompass the over density of stars (shown in Figure \ref{fig:groupXYZ}). The parameters of these ellipses are somewhat arbitrary, and many true members will be more distant. However, inspection of the light curves of stars outside these cuts suggests there are far more interlopers (slow rotators) than true members (stars showing clear young-star variability); the opposite is true inside the ellipses. The ellipse cuts also provided a sufficient list of likely members for a precise age determination. 

Our final membership list includes 125 stars. We list the candidate members in Table \ref{tab:fullList}. In addition to TOI-2076, the membership list includes TOI-1807, which was first seen to be comoving with TOI-2076 in \cite{Hedges2021}.

\section{Update group age}\label{sec:Age}

In order to derive a precise age for TOI-2076's parent population, we combine the age determinations using four methods: gyrochronology, Li depletion, isochronal modeling, and variability-based aging. 

\subsection{Gyrochronology}\label{sec:prot}

As stars age, their rotation slows due to the loss of angular momentum from magnetized stellar winds. This phenomenon implies that rotation rates can serve as indicators of stellar age — a method known as \textit{gyrochronology} \citep{Barnes2007,vanSaders2016}.

Using the \texttt{unpopular} package \citep{2022AJ....163..284H}, we generated \tess\ light curves for the 125 stars in the final membership list. We measured rotation periods using a Lomb-Scargle periodogram \citep{1976Ap&SS..39..447L, 1982ApJ...263..835S} with a linearly spaced search grid of 100,000 steps spanning 0.2 to 20 days. The rotation period corresponding to the highest Lomb-Scargle power is adopted as the measured period. For stars observed in multiple sectors, we ran the Lomb-Scargle on each sector separately and adopted the period from the sector with the highest measured power. Uncertainties in the rotation measurements are estimated using the empirical uncertainty relation from \cite{Boyle2025}.

Given \tess's large 21-arcsecond pixel size, close binary companions may introduce spurious rotation signals into the target star's light curve, potentially leading to incorrect period measurements. To minimize this effect, stars with a Gaia RUWE $>$ 1.2 or flagged as \texttt{non\_single\_star == 1} were excluded.

Each star's light curve was manually inspected to validate the period measurement. Light curves are categorized as ``good'' (a clear rotation signal with a period visually consistent with the Lomb-Scargle result), ``fair'' (a less distinct rotation signal or noisy light curve), ``poor'' (no visible rotation or an evidently incorrect period), or ``binary'' (multiple superimposed signals). To ensure reliable gyrochronological age derivations, only measurements classified as ``good'' are included.

We find the rotation period of TOI-2076 to be $P_{rot} = 7.36\pm0.30$ days with a ``good'' quality. This is consistent with the value found in \citet{Nardiello2022} ($7.29\pm0.12$ days) and \citet{Hedges2021} ($7.27\pm0.23$ days).

The rotation-based age of the group is calculated using \texttt{gyro-interp} \citep{2023ApJ...947L...3B}, which requires the star's rotation period (and its uncertainty) and effective temperature (and its uncertainty) as inputs. Effective temperatures are derived by de-reddening each star's Gaia DR2 $G_{\rm BP} - G_{\rm RP}$ colors using the STILISM dust maps \citep{2019A&A...625A.135L}, followed by conversion to effective temperatures using the empirical color-temperature relation from \cite{2020ApJ...904..140C}.

We generated the age posterior distribution for each star with \texttt{gyro-interp}. To infer the group’s age, the posteriors are combined using \texttt{PosteriorStacker} \citep{2020MNRAS.498.5284B}, which assumes a Gaussian intrinsic age distribution. This procedure yields a final gyrochronology age for the group of $238^{+69}_{-60}$ Myr.

\begin{figure}
    \centering
    \includegraphics[width=0.98\linewidth]{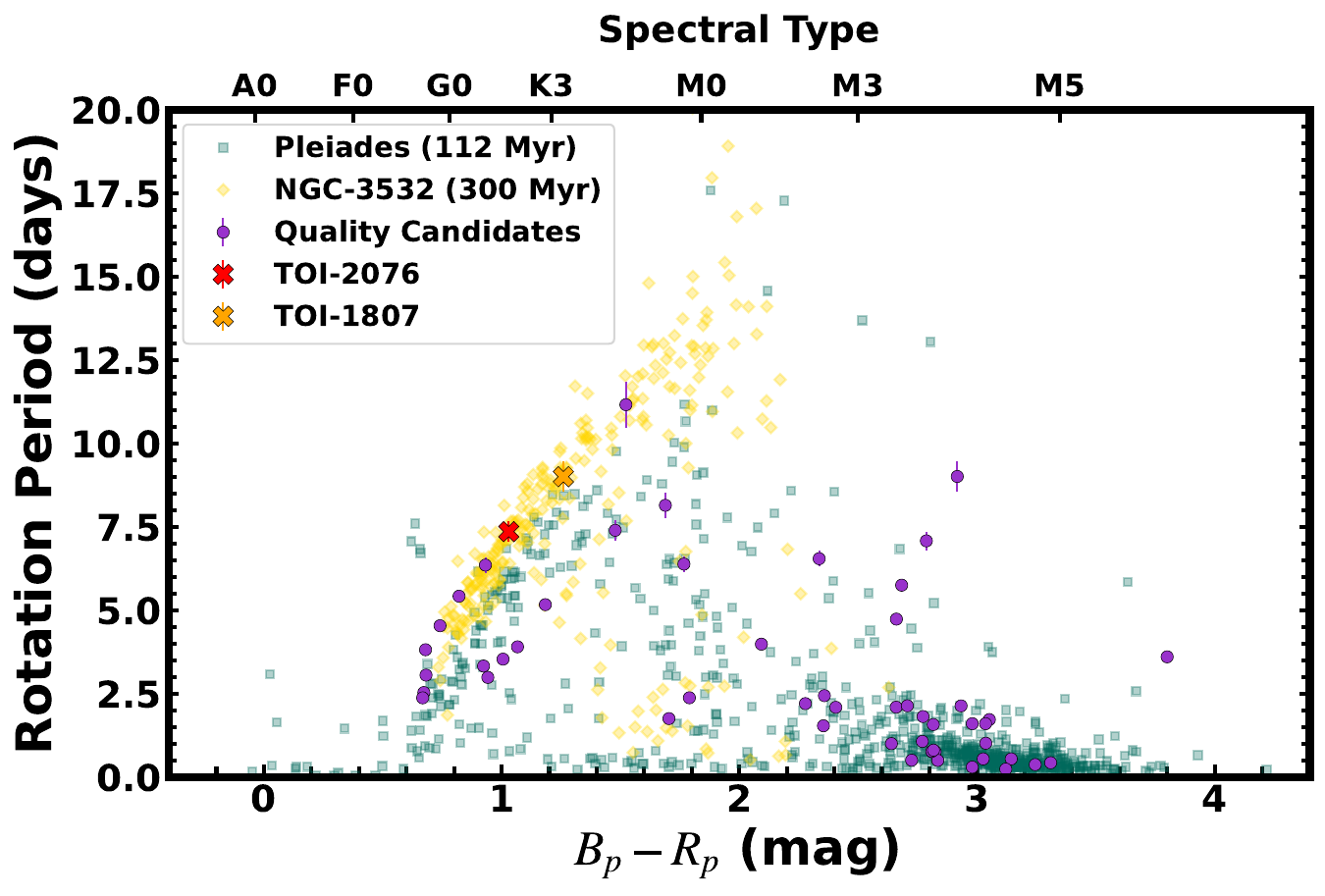}
    \caption{The rotation periods of the candidate co-moving stars (purple circles) against color (as a proxy for stellar type). Only the high-quality, high-reliability rotation periods are shown. TOI-2076 is marked as the red x, and TOI-1807 is marked as the orange x. The rotation sequence for Pleiades \citep[green squares;][]{Rebull2016} and NGC-3532 \citep[yellow diamonds;][]{Fritzewski2021} are shown for reference. The stars co-moving with TOI-2076 show a distribution in between Pleiades and NGC-3532, consistent with the derived gyrochronology age. }
    \label{fig:prot}
\end{figure}

\subsection{Lithium Sequence}

The Lithium (Li) sequence is a method of determining the age of an association based on the observed Li levels of stellar members \citep[e.g.,][]{Jeffries2017, Cummings2017, Wood2023Li}. Li is destroyed in stars with core temperatures greater than $2.5 \times 10^6$ K. Different stellar masses reach this interior temperature at different rates, and mixing can deplete Li levels in the photosphere, leading to a relationship between age and observed Li levels relying on the boundary between stars which have burned their Li and stars that have not \citep[e.g.,][]{Binks2014, Binks2021}. 

We drew Li equivalent width (EW) measurements from \cite{Nardiello2022} for eight stars. Using \texttt{EAGLES} \citep{Jeffries2023}, we estimated the cluster's age based on the Li EW of the candidate co-moving members. \texttt{EAGLES} requires the stars' effective temperatures (rather than color), so we converted the \gaia\ DR2 $G_{\rm BP} - G_{\rm RP}$ measurements to $T_{eff}$ following the same methodology as in Section \ref{sec:prot}.

This resulted in an age of $210^{+45}_{-37}$ Myr.

\begin{figure}
    \centering
    \includegraphics[width=0.98\linewidth]{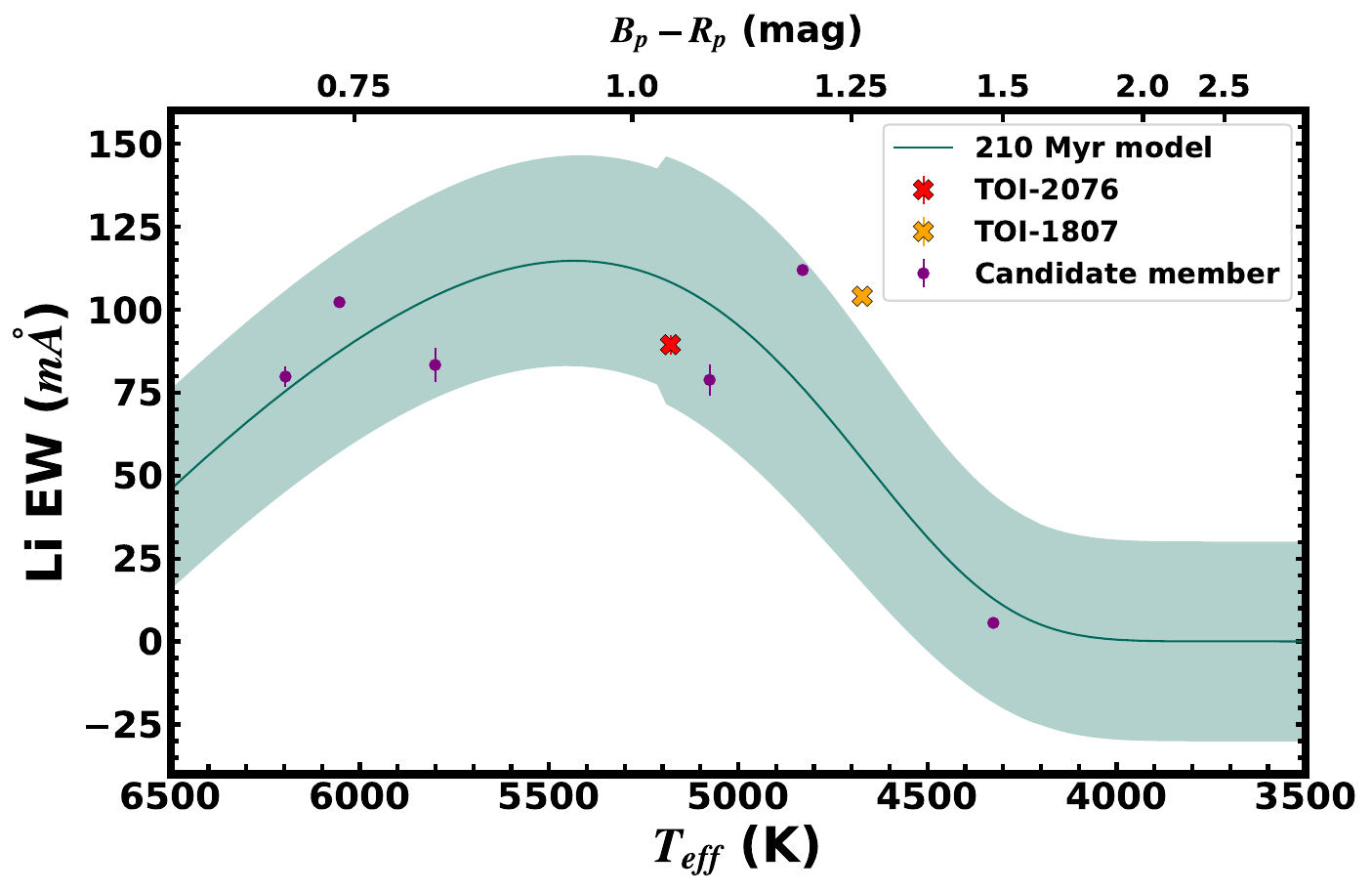}
    \caption{The Li equivalent widths of the candidate co-moving stars against effective temperature (as a proxy for color; purple points). TOI-2076 is shown as the red x, and TOI-1807 is shown as the orange x. The teal region shows the expected distribution for the best-fit age (210 Myr). }
    \label{fig:li}
\end{figure}

\subsection{Isochronal Modeling}

We compared the \gaia\ color-magnitude diagram of our updated membership list to PARSECv1.2 isochrones \citep{PARSEC}. For this comparison, we used a Gaussian mixture model following the appendix of \cite{THYMEVI} and \cite{Wood2023}. To briefly summarize, we used a likelihood formed from the mixture of two components. The first described a single-age single-star sequence described by age ($\tau$) and extinction ($E(B-V)$); metallicity is assumed to be Solar. The second component describes the outliers (e.g., non-members or binaries). Since our membership list is already relatively clean of non-members, we forced the second component to behave like a binary sequence, modeled as an offset from the single-star sequence and a variance ($Y_B$ and $V_B$). Two additional free parameters describe the amplitude of the outlier model ($P_B$) and underestimated uncertainties in the model and/or photometry ($f$). The variables $Y_B$, $V_B$, $f$ are all measured in magnitudes.

We performed the fit using a Markov chain Monte Carlo (MCMC) framework with \texttt{emcee} \citep{emcee}. We ran the fit using 50 walkers for 50,000 steps after an initial burn in of 10,000 steps. Because all stars are within the local bubble, we placed a weak Gaussian prior on extinction ($0\pm0.1$\,mag); all other parameters evolved under uniform priors. 

\begin{figure}
    \centering
    \includegraphics[width=0.98\linewidth]{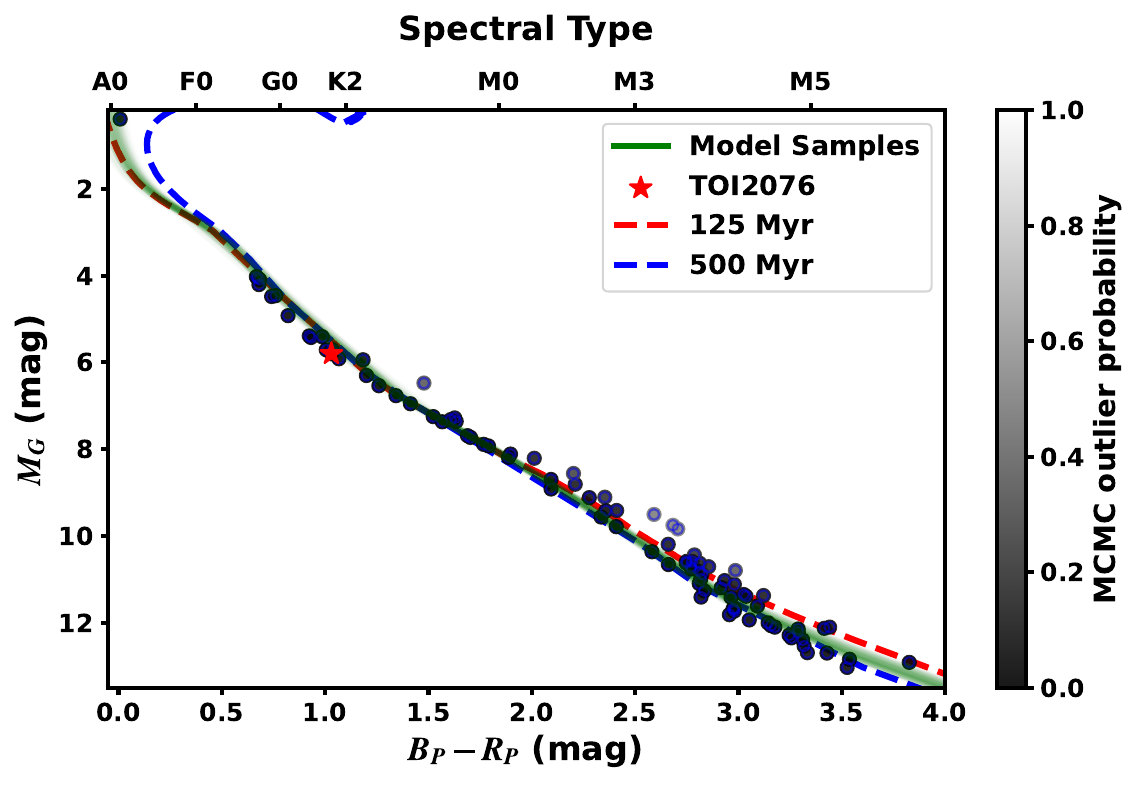}
    \caption{Color-magnitude diagram of likely members of Crius 224 (black circles) including TOI 2076 (red star) compared to isochrones from PARSEC. The green lines are random samples from the MCMC posterior. Circles are colored by their outlier probability (mixture weight). We show 125\,Myr (red dashed) and 500\,Myr (blue dashed) lines for reference. The lack of pre-main-sequence M dwarfs rules out ages $\lesssim$150\,Myr, while higher ages are ruled out by the A star HD 153808. }
    \label{fig:isochrone}
\end{figure}

The final fit gave an age of $\tau = 195\pm21$\,Myr (Figure~\ref{fig:isochrone}). Letting metallicity float within near-Solar values ($-0.3<[M/H]<+0.3$) yielded a less precise but similar $\tau = 197\pm26$\,Myr and still favored a near-Solar metallicity (0.05$\pm$0.10). We adopt the latter as our isochronal age. 

Most groups at this age and mass do not have a clear population of pre-main-sequence stars detected by \gaia, and they are not massive enough to have a large population of turn-off stars. Prior such analyses have yielded a nearly uniform age from 150-800\,Myr \citep{Thao2024}. In this case, the constraints on the upper age end rely mostly on a single A0V star (HD 153808 or $\epsilon$ Her). 

HD 153808 (Gaia DR3 1310284059747409920) has a radial velocity of $-23$ to $-25$\,\kms \citep{Tokovinin2018,GaiaCollaboration2023}, strongly indicating it is a real member (perfect agreement is $-23.4$\,\kms). The star is a spectroscopic binary \citep{Pourbaix2004}, but the CMD is predicted to be nearly vertical in this range (Figure~\ref{fig:isochrone}), so this does not impact the analysis. Prior analyses of HD 153808 have also yielded similar ages \citep[150-230\,Myr;][]{Rieke2005,David2015}. This also agrees with all our other age constraints. Thus, we opt to include the isochronal age in our group age analysis.

\subsection{Variability}

\cite{BarberMann2023} developed a method of estimating ages of young stellar clusters using the excess uncertainties in \gaia\ photometry \citep{Riello2021}. The method takes advantage of more variable sources having higher photometric uncertainties than their quiet counterparts to fit a Skumanich-like relation between stellar activity and age. \texttt{EVA}\footnote{\url{https://github.com/madysonb/EVA}} (Excess Variability-based Age) queries \gaia, makes appropriate stellar cuts, and fits for the age. Inputting the list derived in Section \ref{sec:Age}, \texttt{EVA} calculated an age of $252^{+121}_{-82}$ Myr taking into account all three bands; $G, G_{BP},$ and $G_{RP}$. The variability-based age is robust against field interlopers but the uncertainties are strongly affected by the number of stars in the sample. Given our group is only 125 stars, the large uncertainties are unsurprising.

\subsection{Combining age determinations}

\begin{figure}
    \centering
    \includegraphics[width=0.98\linewidth]{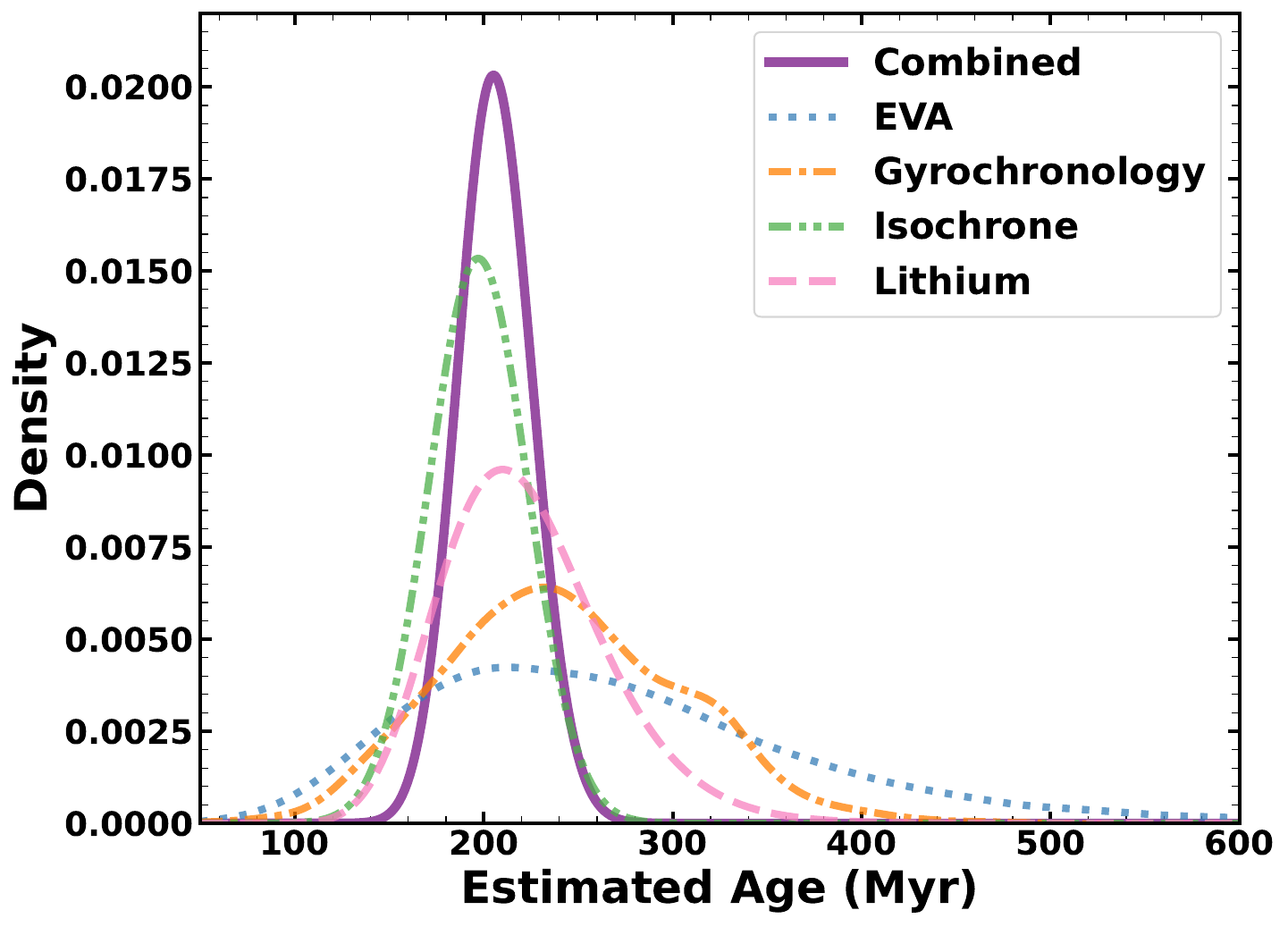}
    \caption{The final age estimate distribution ($210\pm20$ Myr; solid purple) compared to the age estimates from each method (colored lines). }
    \label{fig:combinedAge}
\end{figure}

We combine the individual age determinations for Crius 224 by taking into account the uncertainties calculated in each estimate. The estimates are not truly independent; the gyrochronological, lithium, and variability ages are calibrated using similar clusters and there are known relations between gyrochronological and lithium ages \citep[e.g.,][]{Bouvier2018}. However, each estimate is limited by random uncertainties and the methods contain a mix of model-based methods (isochrone fitting) and empirical ones (gyrochronological and lithium), and rely on different sets of stars (e.g., an isochrone is mostly from the A stars while rotation and lithium mostly comes from FGK stars). Further, even if the absolute age uncertainty is underestimated, the combined age will be robust in a relative sense (compared to the calibration clusters). 

Taking into account each estimate's likelihood distribution (see Figure \ref{fig:combinedAge}), we find an overall age of $210\pm20$ Myr for Crius 224. This estimate more closely agrees with the results of \cite{Hedges2021}, and is consistent with \cite{Damasso2024} to $\simeq1\sigma$ and \cite{Osborn2022} to $2\sigma$. Our value is also far more precise than prior determinations, with uncertainties in line with similar-aged associations \citep{Meibom2011,Thao2024}.

Numerous studies have found evidence of age spreads in young clusters, often in the form of an extended main-sequence turn-off \citep[e.g.,][]{Gossage2019} or a spread in the luminosity of pre-main-sequence stars \citep[e.g.,][]{Cao2022}. Our quoted uncertainty would depend on both measurement uncertainties and any age spread, as such effects would broaden the posteriors for any of the methods above. However, these spreads are usually small \citep[$<5$\,Myr;][]{Jeffries2011,Ratzenbock2023} compared to the 20\,Myr uncertainty here, and there is no evidence of sub-groups in the kinematics of Crius 224. Thus, we attribute most or all of the 20\,Myr uncertainty to measurement and model limitations for measuring the age. Age spreads might be detectable in such groups with significantly more data.

\section{Known planets in Crius 224}\label{sec:knownPlanets}

We searched for additional planet systems in Crius 224 by cross-matching our initial (pre-ellipsoid cut) candidate co-moving stars list against the known planets and TOI list. We identify planet-hosting stars overlapping in the field, some of which may be true members. We discuss individual systems below, organized by detection method.

\subsection{Transiting Systems}

TOI-1807 \citep[TIC 180695581, Gaia DR3 1476485996883837184;][]{Hedges2021} was previously determined to be a member, which we confirm here. The rotation period and lithium are consistent with a similar age, and the six-dimensional coordinates ($XYZUVW$) are within the core of the group (Figure~\ref{fig:groupXYZ}).

TOI-1266 marginally falls outside of our ellipsoidal cuts and exhibits a slow rotation period ($>$20 days). There are no other signs of youth \citep{Stefansson2020}, so we conclude this star is not a member.

\subsection{Directly Imaged Systems}

G 196-3 \citep[Gaia DR3 824017070904063104;][]{Michel2024} is a directly imaged system that lands slightly outside the ellipsoidal bounds in $Z$, and has a modest tangential velocity offset (3.5\kms) from the core of the group. We find a short rotation period ($1.31 \pm 0.01$ days) consistent with Crius 224. However, the host is an M3V; such cool stars can maintain fast rotation out to $\simeq600$\,Myr \citep{2023ApJ...947L...3B}. We consider this a potential, but unlikely member. If G 196-3 is a member, this increases the age. Prior determinations place G 196-3 at $55-135$ Myr by UV emission \citep{Galicher2016} and $10-150$ Myr based on the Li EW of the stellar companion \citep[G 196-3 B;][]{Kirkpatrick2008, Allers2013}. 

\subsection{Radial Velocity Systems}

HD 168746 \citep[Gaia DR3 4153637759337630720;][]{Pepe2002} did not make our list, but was in the original Crius 224 list from \citet{Moranta2022}. HD 98736 \citep[Gaia DR3 3977050728669174912;][]{Ment2018} and HIP 55507 \citep[Gaia DR3 788340461203094912;][]{Feng2022} both make the \texttt{FriendFinder} list. All three systems sit outside the ellipsoidal boundaries and have only marginal evidence of membership. The latter two both have RVs 6--10\kms\ from the core of Crius 224; too far to be explained by random motion or measurement errors. HD 168746 has a rotation period consistent with membership ($9.30 \pm 0.49$ days, $B_p-R_p = 0.86$), but this period is considered unreliable due to having only one sector of data.

\section{Additional Candidate Planets as Candidate Group Members}\label{sec:additionalSearch}

We searched the full list of identified co-moving members for transiting planets. Of the 489 stars in our initial group list, we exacted light curves and searched for planets for 435 stars following the methodology described in sections \ref{sec:extraction} and \ref{sec:search}, adjusting the \texttt{notch} filtering-window based on the rotation period of the star.

No signal was both a clear group member and a clear planetary signal. We discuss individual signals and host stars below. We summarize the results in Table \ref{tab:candidates}.

\begin{table*}[ht]
    \centering
    \begin{tabular}{cccccc}
    \hline
    TIC & Gaia DR3 & $T_0$ & Period & Approx. Depth & Member? \\ 
        &          & (BJD-2457000) & (days) & (\%) &  \\
    \hline
    \multicolumn{6}{c}{Planet Candidates}\\
    \hline
         8673878 & 1324251396473528832 & 1971.46 & 16.01 & 0.5 & Y \\
         17563961 & 1329910067425244160 & 1972.01 & 16.29 & 0.05 & Y \\
         68033619 & 1488431297366254464 & 1744.07 & 26.55$^{\dagger}$ & 0.3 & N \\
         400231203 & 4549900100674763264 & 1984.52 & 27.78 & 0.02 & N \\
                   &                     & 1986.40 & 16.55$^{\dagger}$ & 0.02 & N \\
    \hline
    
    \multicolumn{6}{c}{Other Object Candidates}\\
    \hline
        16271601 & 1278340773059748608 & 2714.21 & 721.38$^{\dagger}$ & 12 & Y \\
        10150705 & 767786671869455360 & 1899.57 & 1.09 & 0.07 & N \\
        392909786 & 3714227557974504576 & 1928.22 & 0.204 & 60 & ? \\
    \hline
    \multicolumn{6}{l}{$\dagger$ likely alias}
    \end{tabular}
    \caption{Candidate signals identified in our initial, uncut list of 489 comoving stars with \starname. }
    
    \label{tab:candidates}
\end{table*}

\subsection{Planet Candidates}

We recover a transit signal in TIC 8673878 (Gaia DR3 1324251396473528832) with a period of 16.01 days, $T_0=1971.46$ TESS-BJD, and depth $\sim0.5\%$. Due to data gaps, this results in only three detected transits, the last one with a flare shortly following egress (poor quality). TIC 8673878 sits inside of our ellipsoidal cuts. It has a rotation period consistent with membership for its color ($P_{rot} = 7.08$ days, $B_p-R_p = 2.79$) and a tangential velocity within 3\kms\ of expected. TIC 8673878 is a likely true member of the group, but the quality of the transits is insufficient to confirm this as a planet.

TIC 17563961 (Gaia DR3 1329910067425244160) is a known wide binary system \citep[2065.1 AU;][]{Tian2020}. Around the primary, we recover a signal with a period of 16.29 days, $T_0 = 1972.01$ TESS-BJD, and depth $\sim0.05\%$. We recover 3 consecutive transits in the first two sectors of \tess\ data available, but transits in the next two sectors all fall in data gaps. Only one signal is recovered in the last sector, and this falls on the edge of a data gap. Both \texttt{FriendFinder} and \citet{Moranta2022} put TIC 17563961 into Crius 224. The star exhibits a fast rotation period ($P_{rot}$=2.37 days) and tangential velocity within 2\kms\ of expected. TIC 17563961 is a likely true member of the group, but the transit is too shallow and has too few transits to confirm.

In TIC 68033619 (Gaia DR3 1488431297366254464), we recover a $\sim$0.3\% signal with $T_0=1744.07$ TESS-BJD and a period of 26.55 days. However, we do not recover consecutive transits so this period may be an alias or a false positive. TIC 68033619 was grouped with TOI 2076 with \texttt{FriendFinder} and sits inside the ellipsoidal bounds. The rotation period is short ($P_{rot}$=2.09 days), but the radial velocity is $>$10\kms\ off from expected. TIC 68033619 is unlikely to be a true member of the group.

We recover a shallow (depth $\sim$0.02\%), 27.78 day signal with $T_0=1984.52$ TESS-BJD in TIC 400231203 (Gaia DR3 4549900100674763264). Only 3 transits are visible across the 5 sectors. We also recover a signal with a similar depth ($\sim$0.02\%), $T_0=1986.40$ TESS-BJD, and a period of 16.55 days, though visual inspections suggest this is likely an alias of the true period. TIC 400231203 was identified as a candidate group member with \texttt{FriendFinder}, though it falls outside of our ellipsoidal cuts. TIC 400231203 exhibits a fast rotation period ($P_{rot}=4.03$ days), though with a poor quality and visual inspection suggests a much longer rotation period. The radial velocity is $>$10\kms\ off from expected for membership. TIC 400231203 is not a likely member of the group.

\subsection{Other object Candidates}

We identify two transit-like events in the light curve of TIC 16271601 (Gaia DR3 1278340773059748608) at 2714.21 and 3435.59 TESS-BJD. Due to the sector breaks, we cannot recover a period for the system. The single transit at 2714.21 was previously reported as a Community TOI \footnote{\url{https://exofop.ipac.caltech.edu/tess/target.php?id=tic16271601}}. The signals are deep ($\sim$12\%) and V-shaped, making it a likely eclipsing binary system. TIC 16271601 falls within the ellipsoidal cuts. The rotation period ($P_{rot}=0.5$ days) is consistent with group membership, making TIC 16271601 a likely true group member.

TIC 10150705 (Gaia DR3 767786671869455360) is a previously reported spectroscopic binary \citep{Pourbaix2004}. We recover a transit-like signal with a period of 1.09 days and $T_0 = 1899.57$ TESS-BJD, consistent with the TCE reported for the system for Sector 48\footnote{\url{https://exo.mast.stsci.edu/exomast_planet.html?planet=TIC10150705S0048S0048TCE1}}. The signal exhibits odd-even differences, consistent with its binary status. TIC 10150705 falls outside the ellipsoidal bounds. The light curve exhibits a rotation period of $P_{rot}=5.64$ days, slightly slow for membership for its color ($B_p-R_p=0.76$), but has a radial velocity $>$10\kms\ from expected, though this may be due to its binary status. TIC 10150705 is unlikely a true member of the group.

TIC 392909786 (Gaia DR3 3714227557974504576) has \tess\ TCE reports with a period of 0.408 days and a period of 0.204 days. It is a known eclipsing binary \citep{Pribulla2003, Prsa2022}, with a true period of 0.204 days. Due to the short period, we do not recover the signal in our pipeline (our minimum search period is 0.5 days). TIC 392909786 falls outside our ellipsoidal bounds. Our rotation period pipeline identified the binary signal as the rotation period, and we were unable to estimate the true rotation period through visual inspection. The close binary orbit also makes radial velocities challenging, with Gaia unable to recover a radial velocity for the star, hindering our ability to confirm or dispute membership. 

\section{Summary and Discussion}\label{sec:summary}

We report the discovery of TOI-2076 e, a super-Earth interior to the three previously discovered sub-Neptune-sized planets in the system. We find TOI-2076 e to be 1.35 R$_\oplus$ on a 3.022 day orbit. We re-derive the age of the system using co-moving stars in the field and previous lists of identified potential cluster members. Using rotations, Lithium levels, isochronal modeling, and variabilities, we find the parent population of TOI-2076 (Crius 224) to be $210\pm20$ Myr and note other planets and planet candidates that likely reside within the association. 

TOI-2076 is a target of the \textit{Keys to Revealing the Origin and Nature Of sub-neptune Systems} (KRONOS) \jwst\ program \citep[GO 5959;][]{Feinstein2024_kronos_prop} focusing on the atmospheric compositions of young sub-Neptunes in multi-planet systems. TOI-2076 is also of interest because of significant TTVs, which may eventually yield masses and eccentricities for the planets \citep{Osborn2022}. The previously unknown planet could effect the interpretation of the system, both the TTVs \citep[e.g.,][]{Weisserman2023} and as part of a comparison between planets in this system. The updated age may also be important for the interpretation of the masses and atmospheres of all four planets. 

TOI-1807 is part of two \jwst\ programs; one searching for silicate vapor atmospheres \citep[Cycle 3, GO 4818;][]{Mansfield2024_jwst_prop}, and one studying the evolution of lava worlds \citep[Cycle 4, GO 8864;][]{Dang2025_jwst_prop}. Similar to TOI-2076, the updated age may be important for interpreting the atmosphere and the evolution of such lava worlds.

As we show in Figure \ref{fig:customVSpdcsap}, our custom light curve extraction is capable of detecting smaller planets than the default extraction using the PDCSAP pipeline. This is a more extreme example than typical, and the SAP flux for the sector does not contain the $\sim0.5$ day spurious signal. However, the PDCSAP flux is commonly used in young planet searches \citep[e.g.,][]{THYMEIV,THYMEVI,Vach2022, Wood2023}, highlighting the importance of light curve extraction in planet search sensitivity. In \cite{Barber2024_hipc}, we were similarly unable to detect the additional transiting planet in the PDCSAP flux to a high enough SNR in our search pipeline despite a strong signal using our custom extraction. These discoveries highlight the need of not only sensitive search pipelines but careful handling of data extraction as well.

The mature \kepler\ multi-planet transiting population displays a higher level of intra-system uniformity in period and radius space than the young \tess\ multi-planet systems \citep{Lissauer2011, Dawson2016, Weiss2018}. With only a limited sample of young, $\ge$2-planet systems, it is unclear if this is true for young planets, especially given detection bias and decreased sensitivity to young planets largely from \ktwo\ and \tess\ systems in comparison to mature planets seen by \kepler. 

The discovery of TOI-2076\,e hints that detection bias is making young systems appear {\it more} uniform. Among the young $\ge$3-planet systems, TOI-2076 was the only system to show a high-level of intra-system uniformity comparable to the mature \kepler\ planets. The detection of the smaller, inner planet pushes the system {\it{out}} of radius and period uniformity.

\begin{figure}
    \centering
    \includegraphics[width=0.99\linewidth]{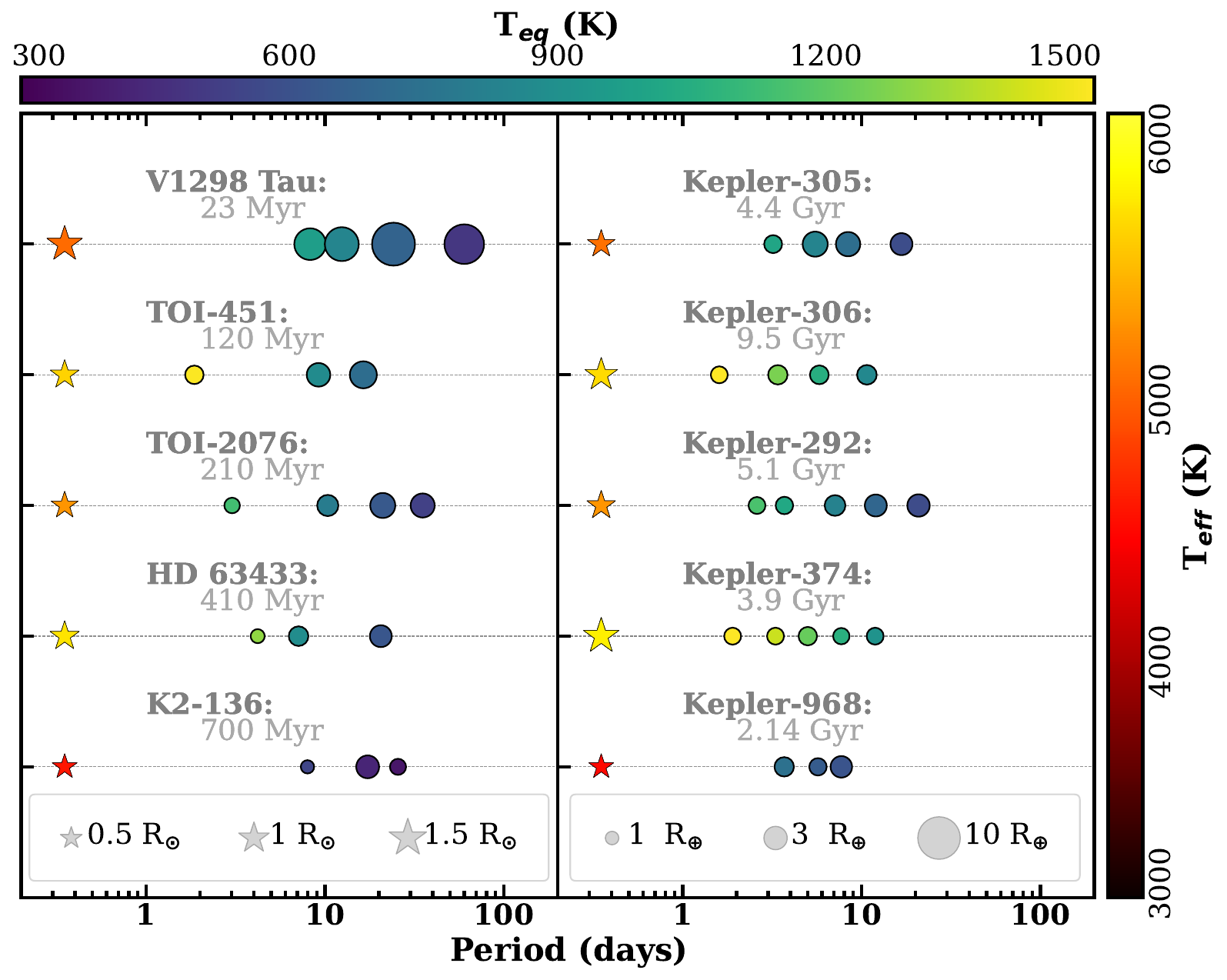}
    \caption{Young, multi-planet transiting systems found in a stellar cluster or association (left) compared to mature multiplanet systems from \kepler\ around similar host stars (right). The mature $\ge$3-planet systems show a higher level of intra-system uniformity in period and radius space than the young systems. The inclusion of TOI-2076 e pushes the system {\it out} of uniformity.}
    \label{fig:uniformity}
\end{figure}

\cite{Dai2024} and \citet{Hamer2024} note an excess of young planetary systems in mean motion resonance (MMR). While TOI-2076 e is not in MMR with neighboring TOI-2076 b (3.4:1), TOI-2076 e is just outside a 7:1 MMR with TOI-2076 d (6.95:1). TOI-2076 b and c (2.03:1) and c and d (1.67:1, or 5.01:3) are near MMR as well. A fifth planet with an orbital period near 6 days would complete the resonance chain 2:1-5:3-2:1-5:3. While our candidate signal at 6.25 days does not pass our quality thresholds, future \tess\ data could be used to search for missing planet f.

The discovery of TOI-2076 e is motivation to revisit previously identified transiting systems to look for additional signals. With a more complete search and a good understanding of our completeness (e.g., from injection/recovery) we can control for biases intrinsic to young stars and map out evolutionary effects like intra-system uniformity. As our survey continues, we expect to find longer period planets due to additional data, but we may also become more sensitive to smaller planets like TOI-2076\,e.

\section*{Acknowledgments}
The authors thank Halee, Bandit, and Elsa for their invaluable support during the creation of the manuscript. M.G.B. was supported by the NSF Graduate Research Fellowship (DGE-2040435) and the \tess\ Guest Investigator Cycle 6 program (22-TESS22-0013). A.W.M. was supported by the NSF CAREER program (AST-2143763) and a grant from NASA's Exoplanet Research Program (XRP 80NSSC21K0393). A.W.B was supported by the NSF Graduate Research Fellowship (DGE-2439854) and funding from NASA's Astrophysics Data Analysis program (ADAP; 80NSSC24K0619). A.I.L.M was supported by the Chancellor's Science Scholars Program\footnote{\url{https://chancellorssciencescholars.unc.edu/}}.

Funding for the \tess\ mission is provided by NASA’s Science Mission Directorate. We acknowledge the use of public \tess\ data from pipelines at the \tess\ Science Office and at the \tess\ Science Processing Operations Center. Resources supporting this work were provided by the NASA High-End Computing (HEC) Program through the NASA Advanced Supercomputing (NAS) Division at Ames Research Center for the production of the SPOC data products. \tess\ data presented in this paper were obtained from the Mikulski Archive for Space Telescopes (MAST) at the Space Telescope Science Institute.

\vspace{5mm}
\facilities{TESS}

\software{\texttt{MISTTBORN}, \texttt{FriendFinder}, \texttt{Notch \& LOCoR}
          }

\bibliography{planetSearch}{}
\bibliographystyle{aasjournal}

\clearpage

\begin{longrotatetable}
\begin{deluxetable*}{cccccccccccccccccc}\label{tab:fullList}
    \tablehead{
    \colhead{Gaia DR3} & \colhead{$\alpha$} & \colhead{$\delta$} & \colhead{Gmag} & \colhead{$B_p$-$R_p$} & \colhead{$V_{off}$} & \colhead{$RV_{exp}$\tablenotemark{a}} & \colhead{$RV$\tablenotemark{b}} & \colhead{$\sigma_{RV}$\tablenotemark{b}} & \colhead{$P_{rot}$} & \colhead{$\sigma_{Prot}$} & \colhead{$Q$\tablenotemark{c}} & \colhead{Li\tablenotemark{d}} & \colhead{$\sigma_{Li}$\tablenotemark{d}} & \colhead{FF\tablenotemark{e}} & \colhead{Crius\tablenotemark{f}} & \colhead{GAPS\tablenotemark{g}} & \colhead{Final\tablenotemark{h}}\\
    \colhead{} & \colhead{(J2016)} & \colhead{(J2016)} & \colhead{(mag)} & \colhead{(mag)}  & \colhead{(km/s)} & \colhead{(km/s)} & \colhead{(km/s)} & \colhead{(km/s)} & \colhead{(days)} & \colhead{(days)} & \colhead{} & \colhead{(m\r{A})} & \colhead{(m\r{A})} & \colhead{} & \colhead{} & \colhead{} & \colhead{}  
    }
        \startdata
        1490845584382687232 &   217.392 &  39.7904 &  8.92 &  1.03 &       0.0 &             0.0 &  -13.08 &          0.15 &      7.36 &             0.3 &                 g &   89.4 &           3.0 &  Y &     Y &    Y &     Y \\
1500434116051163904 &  207.3735 &   41.627 & 16.60 &  4.48 &      0.18 &            0.18 &     ... &           ... &      1.98 &            0.02 &                 p &    ... &           ... &  Y &     N &    N &     Y \\
1426873554041125120 &    247.76 &   54.227 & 14.40 &  2.96 &       0.2 &             0.2 &  -26.13 &          4.47 &      0.74 &             0.0 &                 p &    ... &           ... &  Y &     N &    N &     Y \\
1311701639472895616 &  248.9159 &  29.0931 & 16.04 &  3.44 &      0.36 &            0.36 &     ... &           ... &      2.57 &            0.04 &                 p &    ... &           ... &  Y &     N &    N &     Y \\
1383172330523940096 &  240.5723 &  40.4968 & 12.55 &  2.36 &      0.37 &            0.37 &  -18.28 &          1.62 &      2.44 &            0.03 &                 g &    ... &           ... &  Y &     N &    N &     Y \\
1476485996883837184 &  201.2826 &  38.9224 &  9.68 &  1.26 &      0.38 &            0.38 &   -7.14 &          0.15 &      9.01 &            0.45 &                 f &  104.0 &           1.0 &  Y &     Y &    Y &     Y \\
1370571450529693952 &  234.4453 &  33.1002 & 14.27 &  2.77 &      0.56 &            0.56 &  -17.89 &          2.45 &      1.07 &            0.01 &                 g &    ... &           ... &  Y &     N &    N &     Y \\
1523644978312478976 &  194.6451 &  38.2799 &  9.03 &  1.18 &      0.58 &            0.58 &   -3.34 &          6.85 &      5.17 &            0.15 &                 g &  111.9 &           1.1 &  Y &     N &    Y &     Y \\
\enddata
    \tablecomments{
    \tablenotetext{a}{Expected radial velocity for membership}
    \tablenotetext{b}{From \gaia DR3 \citep{GaiaCollaboration2023}}
    \tablenotetext{c}{Indicates reliability of the rotation period as described in Section \ref{sec:prot}; g = good, f = fair, p = poor, b = binary}
    \tablenotetext{d}{From \cite{Nardiello2022}}
    \tablenotetext{e}{Indicates source was identified as comoving with TOI-2076 with \texttt{FriendFinder}}
    \tablenotetext{f}{Indicates source was identified as comoving with TOI-2076 in Crius 224}
    \tablenotetext{g}{Indicates source was identified as comoving with TOI-2076 in the GAPS survey}
    \tablenotetext{h}{Indicates source was used in the age analysis}
    }
    \end{deluxetable*}
\end{longrotatetable}

\end{document}